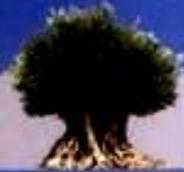

Espace & Développement durable

# VULNÉRABILITÉ ÉQUITÉ ET CRÉATIVITÉ EN MÉDITERRANÉE

Sous la direction de
Yvette Lazzeri et Emmanuelle Moustier

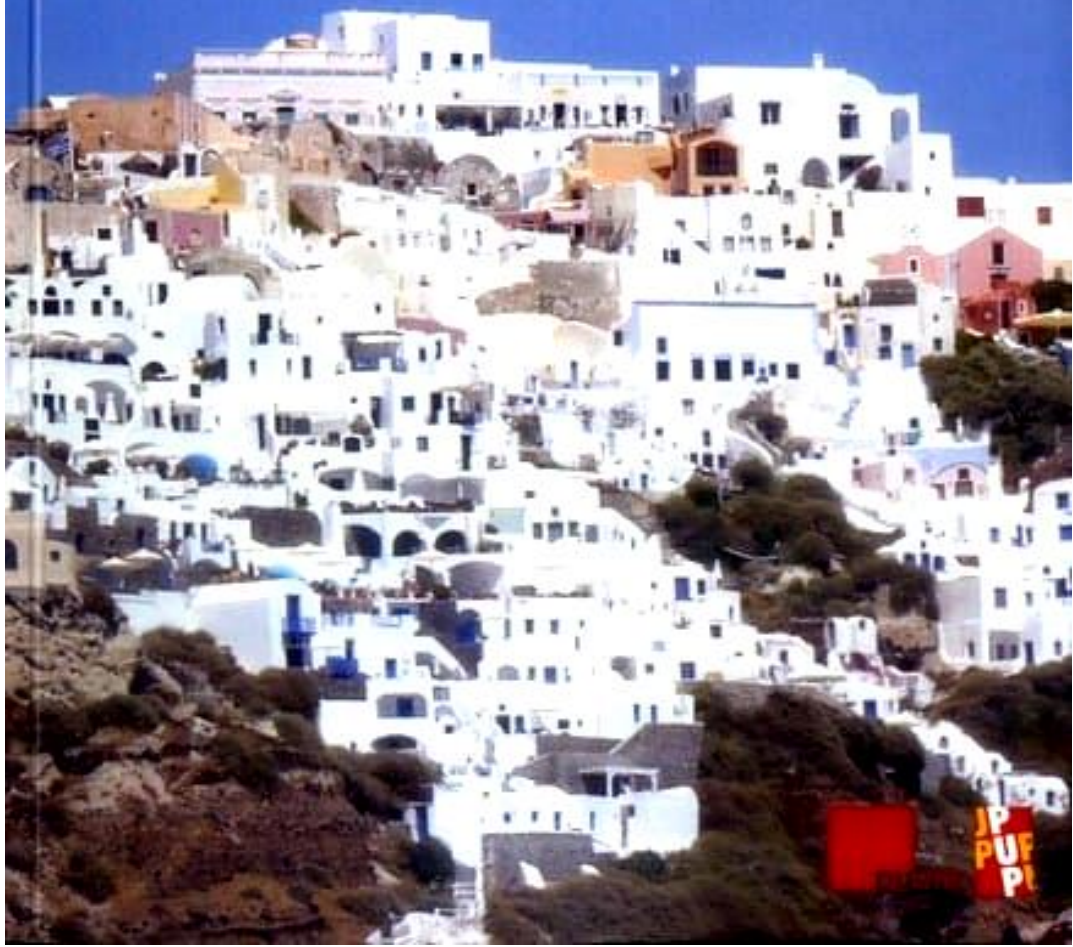

# BARCELONE, LE DISTRICT DE L'INNOVATION 22@ : ENTRE VULNÉRABILITÉ ET ÉQUITÉ SOCIALE

Patrice BALLESTER

## INTRODUCTION : De nouveaux risques, de nouvelles réponses ?

Poblenou, ou le nouveau village (en catalan), est un quartier de la commune de Barcelone connaissant des transformations profondes depuis plus d'une dizaine d'années. L'objectif de la municipalité est de favoriser l'implantation d'industries créatives tout en créant de nouveaux liens sociaux sur un espace en crise au début des années 1990. Un paysage urbain de grandes tours tertiaires issues des ambitions architecturales de sociétés internationales se propose aux habitants de la capitale catalane. Une spéculation immobilière y est fortement présente, ainsi que la confrontation de deux visions territoriales sur cet espace péricentral. Pour les uns : territoire d'entrepôts, studios d'artistes, d'habitats populaires et de grandes friches industrielles à sauvegarder, pour les autres : territoire hautement spéculatif et stratégique de la nouvelle économie dont l'ambiance urbaine se caractérise par l'implantation de gratte-ciel d'architectes reconnus comme la Tour Agbar de Jean Nouvel.
Poblenou ou *«la Manchester du Sud de l'Europe»*[1] est la plus grande friche industrielle d'Espagne. La recherche d'une nouvelle croissance économique et la volonté d'infléchir la destinée territoriale de ce quartier en crise sociale sont au cœur du débat public. Cette volonté se lit dans le règlement du Plan d'Occupation des Sols où les aménageurs font passer la zone 22A de Poblenou (Plan Général Métropolitain des années 1970/1980 – A pour zone industrielle) en zone 22@ au début des années 2000 (PGM révisé à de multiples reprises depuis – @ pour zone de la nouvelle économie). Le A devenant @ pour marquer le remplacement de l'emprise au sol de l'industrie lourde par l'industrie de la connaissance et du numérique. L'autre symbole est de rappeler que le 2 et le @ sur le clavier espagnol sont sur la même touche. La logique de réalisation du projet 22@ repose sur le e-commerce, le high-tech industriel et les talents individuels, voire plus illusoire, la recherche de la sérendipité par une alliance formelle entre *science city*, *cultural city*, *creative city*. L'ensemble s'apparente à un basculement territorial et à un questionnement du devenir de cette ville méditerranéenne et de son urbanité renouvelée dans un *barri* (quartier) historique et symbolique (Andreu, Gol, Recio, 2001).
Dans une ville pensée comme étant dense, compacte, attractive, créative et durable (Busquets, 2004), cette création d'un district de l'innovation bénéficie-t-elle à l'ensemble de la société ? De la vulnérabilité économique par la perte d'emplois industriels, le chômage de masse et l'affaiblissement des budgets municipaux, est apparue une contre-réaction dans l'accompagnement de la mondialisation par un processus territorial original s'apparentant par mimétisme à la *Silicon Valley* en Californie ou au quartier de Chelsea à New York.

---

[1] Nom donné à Poblenou depuis les années 1900, montrant l'importance pour l'Espagne et la Catalogne de cet espace industriel unique dans la péninsule ibérique de par sa dimension (200 hectares) et sa capacité de production en direction des marchés européens.

Les références aux thèses critiquables de Richard Florida (Florida, 2002, 2005) et Michael Porter (Porter, 1998) sont revendiquées par les acteurs du projet dont la personne même de Miquel Barceló, directeur de la structure 22@, au tout début de sa conceptualisation (Barceló, 2001, Barceló, 2009 :76). Pour répondre à cette question, notre démarche se structure en trois interrogations reposant à la fois sur la théorie et la pratique-observation d'un terrain d'étude parcouru à plusieurs reprises entre les années 2004/2011. *Première interrogation* : de la Cité Digitale (Pacte Industriel de la Région Métropolitaine, 2001), ouvrage genèse commandé par la mairie pour penser le territoire de la cité en 2020, au district de l'innovation 22@ loué à l'exposition universelle de Shanghai en 2010 - pavillon de l'Espagne - par quels appareillages les acteurs publics du projet ont pu positionner leurs intérêts mercantiles et fonciers avec les aspirations des habitants (1.) ? *Deuxième interrogation* : Comment par la technicité et l'ingénierie d'une ville durable, l'attractivité des firmes multinationales est recherchée créant de fait un paysage urbain symbolique de grandes tours (2.) ? *Troisième interrogation* : Quelle méthode pour évaluer la réalisation d'un district de l'innovation dans le monde méditerranéen ? Les aménagements de l'espace urbain répondent-ils à la demande initiale de résilience et de croissance économique pour toutes les catégories sociales en presque 20 ans de régénération depuis les JO de 1992 ? Une approche quantitative et qualitative amène à un bilan contrasté dans la recherche d'une équité sociale (3.)

## 1. LES CONDITIONS D'UNE RÉGÉNÉRATION URBAINE PAR L'INNOVATION SOCIÉTALE ET l'ECO-URBANISME ? DE LA VULNÉRABILITÉ À LA CRÉATIVITÉ

Le pragmatisme catalan et l'histoire du quartier de Poblenou expliquent la nature des aménagements conséquents du quartier pour permettre l'attractivité des entreprises de l'innovation (1.1), la créativité doit permettre de proposer un nouveau modèle de ville dans la dotation d'aménités favorables à son développement et insertion dans une compétition mondiale (1.2).

### *1.1 Les mutations de Poblenou : les 4 étapes de la requalification urbaine*

a) *Un plan d'urbanisme ambitieux et gentrification programmée, 2000/2004.* En prenant en compte l'histoire urbaine du quartier et ses impératifs économiques et sociétaux (Solà-Moralés, 1974), la municipalité de Barcelone, détentrice des 100 % de la société d'aménagement et de promotion de Poblenou - arroba 22 s.a.u. - veut utiliser la culture et les activités créatives comme nouveau programme de développement économique en termes d'attractivité et de présentation au grand public d'un milieu innovant. Le cluster doit permettre à la fois croissance économique et requalification urbaine en se fixant comme objectif de se spécialiser dans les entreprises de design, de création graphique, multimédia et TIC médical (Mairie de Barcelone, 2000, Oliva, 2003).
b) *Arrivée d'industries créatives et début de polémique : 2004/2005.* C'est à partir de 2004/2005 que l'arrivée à Poblenou des premières industries créatives se concrétise. Dans un premier temps, les domaines du multimédia, de la publicité et de la télévision sont privilégiés. Une cité néo tertiaire voit un nouvel emblème urbain : la Tour Agbar de Jean Nouvel, pour une société de gestion des eaux. Elle capte l'attention des concitoyens et des décideurs économiques. C'est une nouvelle ligne d'horizon selon le chef de l'Agence d'urbanisme et de planification de Barcelone (Acebillo, 2004). C'est aussi sur le plan du marketing urbain une des toutes premières réalisations attirant l'attention de tous les investisseurs. Une tour qui par essence de jour comme de nuit est immanquablement un objet d'art et de créativité au pouvoir d'évocation et de rêveries par ses illuminations colorisées sur sa paroi grâce à un système astucieux de jeux de lumière Led. L'office du Tourisme de Barcelone a pour objectif d'en faire un nouveau référent identitaire de la ville touristique, mais aussi de la cité des congrès et centres d'affaires faisant reposer sa nouvelle croissance économique sur un district de l'innovation.

c) *Un combat symbolique pour faire changer l'opinion publique 2005/2006.* Pourtant, la période du début de régénération urbaine se situe dans un moment de spéculation immobilière et d'inflation du prix du m² en Espagne. Cette troisième étape se traduit par l'épisode inattendu, mais révélateur des tensions urbaines dans la lutte d'artistes pour une autre gestion alternative de ce territoire provoquant par ricochet la campagne de sauvegarde de Can Ricart, une ancienne usine-forge à la valeur patrimoniale certaine. L'action devient le point focal de la contestation artistique et d'une partie de la population de Poblenou. Le projet est considéré comme une provocation dont les premières manifestations anti Forum Universel des Cultures en 2004 furent les prémisses.

d) *La création de nouveaux espaces publics issus de la mondialisation. 2007/2011.* Nous sommes particulièrement concernés par cette étape, car c'est celle que Poblenou vit actuellement dans un jeu complexe de relocalisations et interactions spatiales en lien avec une grave crise hypothécaire. Des industries créatives comme Google, des hôtels quatre étoiles comme la chaîne Novotel, des lounges-bars, clubs branchés, restaurants et galeries d'expositions reconnues, s'installent à Poblenou. Le secteur est de plus en plus conforme à ce que les planificateurs envisageaient (Vera Martín, Pallarés Barberá, Tulla Pujol, 2008). Face aux combats des artistes, une réaction émane avec une nouvelle légalisation proposant un quadruplement des zones de protection du patrimoine industriel. Le prix du m² dans ce quartier de Barcelone ne connaît pas de pause, même en début de crise économique (La Vanguardia, 2009).

### *1.2 Un laboratoire social, culturel, technologique et créatif*

Les attentes sont grandes concernant un projet dit processus par ces acteurs. Les Jeux olympiques de 1992 passés, la Mairie de Barcelone amorce une seconde renaissance (Masboungi, 1998) devant permettre à la cité de devenir un *«laboratoire urbain et modèle de ville méditerranéenne»* (Montaner, 2006). Elle décide, pour des raisons de crise budgétaire, d'associer le secteur privé au secteur public pour la construction et la régénération de nouveaux quartiers péricentraux comme Poblenou (Mairie de Barcelone, 1998, 1999, 2000 - Busquets, 2004 : 411-444).

La mairie entreprend son action sur le concept de ville durable pour aboutir à un renouvellement urbain optimum *«plus une ville sera complexe, et plus elle sera durable»* : selon les porteurs du projet manquant parfois de recul dans des séances de communication grand public (Broggi, 2007). La ville doit se densifier tout en se diversifiant selon les théoriciens de la nouvelle économie proposant une rhétorique basique sur la ville dite durable. L'ensemble de cette planification n'est qu'une énième utopie (Laigle, 2009) comme le sont aussi les autres dénominations additives au projet : ville verte, ville créative. Elle doit préserver la cité et ses habitants des maux de la mondialisation, de la délocalisation et autoriser à juste titre la relocalisation d'industries de pointe en travaillant sur la notion de créativité pour juguler le chômage de masse (Mairie de Barcelone, 2003).

La municipalité propose dans les premiers temps du projet d'agir en fonction des qualités géographiques et culturelles reconnues de la ville. Le parcellaire de la ville, la grille de Cerdá, et trois autres points névralgiques de la cité méditerranéenne forment un territoire par un triangle : la grande avenue Diagonale percée récemment, se prolongeant jusqu'à la mer et l'espace Forum, le rond-point des Gloires Catalanes avec la Tour Agbar de Jean Nouvel comme emblème architectural mondial et la façade littorale de Poblenou avec ses plages olympiques. Le principe est d'améliorer l'efficience urbaine définie par le rapport ressources/complexité ou diversité des activités. La compacité est l'une des formes urbaines favorisant la complexité dans un espace-temps donné avec l'intensification de l'usage du sol et le respect de la diversité sociale. L'objectif est de créer 150 000 emplois et d'attirer plus de 4500 entreprises horizon 2020[2]. Pour accueillir ces parcs d'entreprises, les aménageurs décident de limiter l'usage de la voiture en agrandissant les trottoirs et démultipliant les services de proximité. On élargit les pistes cyclables et développe d'astucieux systèmes contre le bruit provenant de grandes avenues tout en utilisant les jardins et espaces publics comme des lieux de vie et de rencontres repoussant au loin la route. Il est clairement spécifié dans le règlement d'urbanisme de voir proposer par les agents immobiliers des surfaces habitables de 175 m² par foyer (en moyenne) incluant en plus un réseau de contrôle et de gestion du système de distribution de l'électricité - système *smart grid* par des mises en réseaux et compensation des différents réseaux de production de l'énergie entre le chauffage par biomasse, le solaire, l'éolien et le réseau classique d'électricité. La gestion n'est pas globale (à l'échelle du pays/métropole), mais locale par la gestion de pôles de quartiers ; l'ensemble visant à optimiser la distribution d'énergie sur un principe technique allemand dans sa conception et gestion des risques énergétiques. La construction de plus 4500 immeubles d'habitations sont planifiés sur le long terme.

L'objectif du plan habitat est triple : maintien de loyers modérés, assurance d'une mixité sociale au sein du quartier et possibilité de créer quelques micro-quartiers résidentiels permettant d'édulcorer l'image du quartier par un regard nouveau d'une population aisée. Quinze pour cent de la superficie totale est dédiée à l'habitat dit populaire.

---

[2] On notera qu'une partie du quartier de Poblenou se porte au-delà de ce triangle, à la limite d'un autre quartier, Nou Barri, nommé les franges de Poblenou. Au plan administratif, on parle aussi du district de San Martí.

L'ambition est d'assumer la densité urbaine et d'aménager un cadre de vie confortable avec des espaces publics de qualité en référence aux oasis et belvédères du premier projet urbain des années 1980. Dès lors, on comprend que c'est à partir de l'îlot Cerdá que l'on doit réaménager le quartier. Les formes d'incorporation des projets dans l'îlot, sont démultipliées les îlots sont soudés, divisés ou tout simplement créés (Barcelona Projecta, 2007). On parle de super îlots pour le cluster MediaTic jouxtant le rond-point des Gloires Catalanes et incorporant la tour de Jean Nouvel. Par cette expérience, les urbanistes se démarquent de Cerdá.

La *«Barcelone néo tertiaire»* (Acebillo, 2006, conversation avec Marco Casamonti) a pour objectif le développement économique du quartier dans un rayonnement régional et européen (Sánchez, J.E., 2002). Pour comprendre ce district de l'innovation il faut le rapporter au nouveau modèle métropolitain barcelonais fondé sur l'idée que *«la ville durable est une ville compacte à forte densité spatiale présentant les caractéristiques de complexité fonctionnelle, connectivité, accessibilité, diversité des populations dans l'efficacité du métabolisme urbain et de la cohésion sociale»* (Rueda Palenzuela, S., 2007). L'expression *«systema entorn»*[3] ou *système environnemental rural/urbain* devient slogan et force de loi et de principe du nouveau projet urbain : *«Observer les interactions entre le développement urbain et l'environnement, l'un influence l'autre et vice-versa, le rapport entre ces deux facteurs détermine le degré de durabilité d'une ville. L'objectif est de renverser les données de l'équation en misant sur la complexification des modes de gestion urbaine pour assurer une amélioration du cadre de vie et la préservation des ressources naturelles par un système que l'on appelle le degré/indicateur de compacité d'une ville»* (Tugayé, 2009). Le degré de compacité urbain doit amoindrir les risques urbains de dépenses d'énergie pour se rendre au travail, augmenter la qualité de vie en profitant à pied, si possible, de la ville et de ses espaces publics de qualité, il est en conséquence impératif de proposer : une diversité des fonctions urbaines + l'accessibilité aux transports collectifs + la proximité des services + l'accès à l'espace et aux équipements publics. Or, la localisation des friches industrielles et des locaux nécessaires aux industries innovantes sont un atout et un handicap à la fois, du fait des surfaces à traiter, à entretenir, à dépolluer (autre risque) et à se partager/s'accaparer entre la mairie et les entreprises internationales. Un véritable pari par une grande ambition, mais ne considérant pas certaines réalités physiques et sociales du terrain, ainsi que le boom immobilier espagnol et le changement de structures et d'ambitions des sociétés de promotion immobilières catalanes incluant leurs actions sur une rentabilité maximum et des projets de plus en plus démesurés face à la solvabilité et capacité de crédit des acheteurs de plus en plus précaires (Sánchez, J.E., 2003).

Il faut préserver les équilibres urbains garants de ceux que les Barcelonais appellent le fameux droit à l'urbanité (une démocratie locale renforcée, des espaces publics de qualité, une mixité sociale et croissance économique pour tous). Un paysage urbain de la mondialisation par des figures architecturales et style de vie parfois en complet décalage avec la culture locale se généralise.

---

[3] Le *système environnemental* : l'agence de l'écologie urbaine de Barcelone a pour but de comprendre, analyser et anticiper la ville du futur. < http://www.bcnecologia.net > au 01/07/10

Un quartier durable dans une ambiance innovante bouleverse la vie des habitants, des artistes et des entrepreneurs.

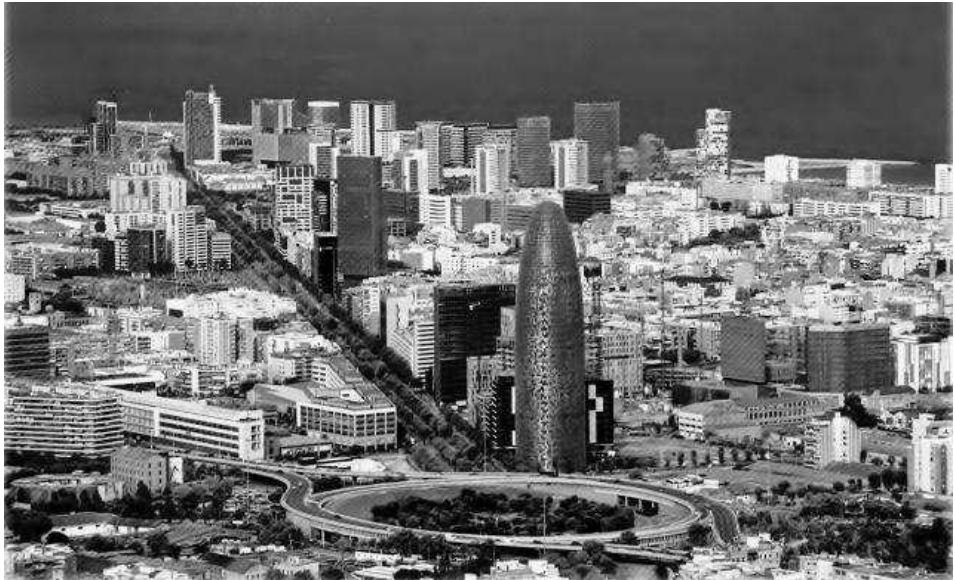

Quartier 22@, 2010, Mairie de Barcelone.

## 2. ATTRACTIVITÉ, DURABILITÉ ET ACTEURS SOCIAUX DU 22@

Les conditions d'attractivité et de durabilité du projet résident dans le vademecum ou cahier des charges souvent pléthore, mais reposant sur une idée d'échange de terrain contre services à la collectivité pour mutualiser les risques avec l'aide de la créativité et de bâtiments écologiques pour y parvenir (2.1). Un projet ambitieux, voire fécond, avec une profusion de conditions et de règlements fonciers pour inspirer, accompagner et surtout distribuer l'usage des sols et les compétences territoriales (2.2).

### 2.1 Les conditions d'une attractivité

La ville recherche donc la synergie, l'émulation, l'attractivité : mots magiques maintes fois décriés par les spécialistes pour les clusters ou pôles de compétitivité avec les échecs inhérents de croire en une économie cognitive fondatrice de nouvelle valeur. Dans ce cas précis, ce projet offre une dimension métropolitaine et régionale de mise en réseau à l'échelle de la Catalogne et de l'Europe sur le modèle américain (Yigitcanlar, Velibeyoglu, Baum, 2008). Il est nécessaire de réunir les activités du design, des médias, des TIC, des technologies médicales et de l'énergie pour enrichir économiquement le quartier tout en profitant de la créativité barcelonaise, des initiatives locales, voire des artistes informels du quartier qui, il faut le souligner, ne sont jamais cités dans les premiers plans de l'agence municipale Arroba 22 au profit de l'association des habitants de Poblenou toujours en veille de surveillance critique (Borja 2004

- El Poblenou 2001). D'autre part, la municipalité renforce la capacité d'innovation des entreprises par la création d'un incubateur unique au pouvoir de décision important notamment dans la capacité de trouver un local et un financement à la mesure du projet. De plus, les quatre clusters sont localisés et référencés précisément sur le territoire de Poblenou. Toute entreprise voulant s'implanter doit respecter le fait de construire son siège dans la zone identifiée comme ouverte à la nature principale de son développement. En outre, la municipalité décide d'y insérer des services publics de l'enseignement supérieur, des universités et écoles de design et de communication. Elle renforce par la même la concentration des centres de R&D et des supposés transferts de technologie favorisant l'attrait de projets d'entrepreneurs les plus innovateurs au niveau international. Ce procédé s'enracine dans la stratégie générale d'aménagement de secteur urbain en essor de la ville où sont concentrées actuellement les opérations les plus importantes.

Trois aires de nouvelles centralités accompagnent le projet 22@ : Sant Andreu Sagrera pivotant autour du projet de la nouvelle gare intermodale de la Sagrera où arrivera le Train à grande vitesse, deuxièmement l'aménagement de la place des Gloires Catalanes, futur centre culturel et administratif, ainsi que lieu du cluster Media-Tic - et enfin le nouvel espace Forum et ses infrastructures de Congrès issues du Forum Universel des Cultures de 2004. On constate que le quartier n'est pas isolé, mais en lien direct avec les autres aires de nouvelles centralités et le reste de l'Europe.

### *2.2 Les conditions d'une durabilité : la technicité et ingénierie pour une écocité flexible et créative*

Les opérateurs ont décortiqué une série de mécanismes et de thèmes principaux pour développer leurs idées et incorporer l'ensemble des actions en les rendant interdépendantes les unes des autres ; espaces publics, nouveaux logements, équipements publics et densité urbaine sont les principaux axes de réflexion. L'ensemble doit encourager la présence d'activités denses de la connaissance dans des infrastructures-bâtiments durables (pas forcément pour tous et avec des polémiques sur les labels) permettant à la fois des avancées dans la mixité sociale avec une donnée clef : la flexibilité. Sept propositions novatrices incitent à croire que ce projet est assurément une originalité dans les territoires créatifs méditerranéens puisqu'on propose de se servir du sous-sol pour permettre de gérer le présent et d'anticiper l'avenir avec la mise en place d'un réseau souterrain complexe, permettant néanmoins de gagner de l'espace, de produire et conduire l'énergie, tout en offrant à tous les citoyens un système d'assainissement faisant de chacun d'eux un acteur énergétique en matière de salubrité publique (source Arroba 22, s.a.u. 2006). Tout d'abord, on propose un service de gaz et d'eau plus performants et efficaces grâce à un système de vannes modernes, puis un nouveau réseau de fibres optiques de télécommunications, se rajoute un nouveau système de climatisation publique, centralisée, afin d'obtenir plus de 35% d'augmentation énergétique disponible par rapport aux systèmes traditionnels. Les très grandes galeries souterraines bétonnées, accessibles, permettent de réparer et d'améliorer les réseaux de service sans faire de travaux habituels de crevasse imposante sur la voie publique. À cela se superpose un nouveau réseau de transports publics, ainsi que la création d'un large réseau de pistes cyclables (29 km) favorisant la fluidité de la circulation pour justement garantir la disponibilité de places de parking pour les travailleurs et les visiteurs du secteur. Le nouveau réseau électrique garantit la qualité d'approvisionnement, moyennant une puissance cinq fois supérieure à l'initiale, permettant ainsi au système de recueil pneumatique sélectif de déchets de fonctionner par énergie/soupape compressée, et en outre, d'éliminer les containers alignés dans les rues (paysage urbain amélioré). Cela supprime également les nuisances sonores et les dérangements de camions poubelles, et enfin assure une rapidité et un tri sélectif dans la distinction de trois éléments : organique, papier et autres. Des règles en matières foncières sont également formulées par les artisans du 22@. La maîtrise totale du foncier est dévolue à la mairie et aux agents de la société 22@. Pourtant, l'initiative privée est à l'origine de la modélisation du nouveau quartier. Les propriétaires d'îlots doivent rencontrer des promoteurs immobiliers pour concevoir ensemble un projet de rénovation urbaine, ils négocent dans un rapport donnant/donnant, l'un apporte le terrain, l'autre le savoir-faire du constructeur (avec très souvent des dons d'appartements au propriétaire du terrain). Puis, ils approchent les services d'urbanisme et siège 22@. S'engage alors une deuxième négociation sur le plan de masse et la hauteur des bâtiments notamment par le coefficient d'occupation du sol à appliquer. Si le projet prévoit de mettre en place une partie de logements sociaux et en plus une implantation d'entreprise reposant *«sur l'économie innovante »* (Ayuntamiento de Barcelona, 1999), on augmente la densité par une autorisation supplémentaire du nombre d'étage. L'aménagement se conçoit à partir des nouveaux îlots Cerdá ; la ville et son foncier sont un objet de croissance innovante, c'est une future source de confrontation avec les artistes et la population locale (Walliser, 2004, Lefevre, Romera, 2007).

La concertation par l'innovation, la connaissance par la législation et le marchandage foncier doivent concilier l'économisme et le juridisme ainsi que la prise de risque de toute la société catalane se projetant dans cette ambition urbaine avec l'aide importante des sociétés privées. La créativité est au rendez-vous et l'exemple de l'urbanisme souterrain au secours de l'équité urbaine se doit d'être souligné. Un autre facteur explique aussi le succès de cette entreprise, la flexibilité : quand un problème survient, on n'hésite pas à changer de législation, soit le curseur s'oriente à la baisse, soit il s'oriente à la hausse, suivant le pragmatisme somme toute catalan. C'est le cas de la récolte pneumatique, pour l'instant non généralisée à Barcelone et mise en sommeil vu les difficultés techniques.

Un bilan est maintenant possible prenant en compte les dernières données chiffrables sur l'attractivité territoriale et la mise en place d'une nouvelle Icarie ou société urbaine issue de l'innovation (Icarie : nom donné à un quartier, partie intégrante de Poblenou, par l'ingénieur Cerdá à la fin du XIXe siècle, symbole de la ville équitable et socialiste).

## 3. LA DIFFICILE ÉVALUATION D'UN DISTRICT CRÉATIF

L'équité, ou *«notion de la justice naturelle dans l'appréciation de ce qui est dû à chacun ; vertu qui consiste à régler sa conduite sur le sentiment naturel du juste et de l'injuste »* (Dictionnaire Robert, 1984) tend à nous faire admettre qu'il faut soi-même tracer la frontière entre ce qui est juste ou injuste.
Il faut prendre en compte une inégalité tronquée due aux interactions sociétales, environnementales et de prise en compte des dettes étatiques ne permettant plus de garantir l'égalité. C'est une question de sémantique, mais aussi pour nous de prise en considération des réactions des habitants, du partage des richesses, de leurs redistributions ainsi que des connaissances statistiques, sans omettre les systèmes de représentation complexes mis en place et annonciateurs de nouvelles valeurs d'usage. Un bilan comptable s'impose et fait parfois rêver si l'on compare cette cité avec l'attraction d'autres villes méditerranéennes comme Marseille ou Gênes (3.1). Pourtant le combat de Can Ricart et les paroles libérées des habitants montrent les limites d'une telle planification (3.2). Des coopérations originales essayent de tisser des liens plus solides entre habitants déboussolés et entrepreneurs sûrs de leur droit (3.2).

### 3.1 Innovation et chiffres : des résultats probants mais paradoxaux

Concrètement, en dix ans d'action, 1502 entreprises se sont implantées à Poblenou, ainsi que 12 centres de recherche et de développement. Onze universités et écoles privées se sont ouvertes, dont certaines sur la thématique du design, de l'art, de la communication visuelle et du webdesign (Chiffre août 2010, Arroba s.a.u.). L'ensemble a permis de créer 44600 postes de travail de haute qualification. Il s'agit non plus d'un territoire pauvre, mais riche, car en dix ans de mutation le volume des transactions s'élève à plus de 6 milliards d'euros. Sur le total des entreprises installées, 74,2 % sont en lien direct avec la nouvelle économie représentant une part de 58% du total des entreprises de la connaissance pour la Catalogne. Sur les cinq clusters/thématiques de développement, on comptabilise : 26 % des entreprises dans le domaine des nouvelles technologies et de la communication, 24 % dans le domaine du design, 11% dans les médias de communication, et 5% dans la technologie médicale et 4 % dans l'énergie, deux secteurs en difficulté. Malgré la crise que connaît l'Espagne, 67% des dirigeants de 22@ estiment que la situation est satisfaisante dans leur secteur et qu'elle ne va pas s'altérer. Enfin, National Geography Europe, CMT et Bassat Ogilvy font partie des 90 dernières entreprises ayant pris position à Barcelone à partir du début 2010. Depuis peu, la mairie vante ce district comme *«modèle de ville»* par la collaboration du public/privé des opérateurs et de *«ville de la diversité, de la cohérence sociale et du développement durable où se rencontrent les transferts de technologies, commerces, habitations, équipements et zones vertes »* (22 Arroba, 2010, interview). Or, quelle est la part d'emplois des habitants de Poblenou dans ces nouvelles industries ? Elle est minime : moins de 5% du total des nouveaux postes, avec souvent des tâches subalternes et réplétives n'amenant rien de créatif. C'est une constante en Europe de rencontrer ce phénomène.

Il reste qu'en 2008, le Poblenou a totalisé davantage de créations d'emplois qu'à l'époque de l'apogée de la zone industrielle, il y a quarante ans de ça, au plus fort de son activité, soit 35 000.

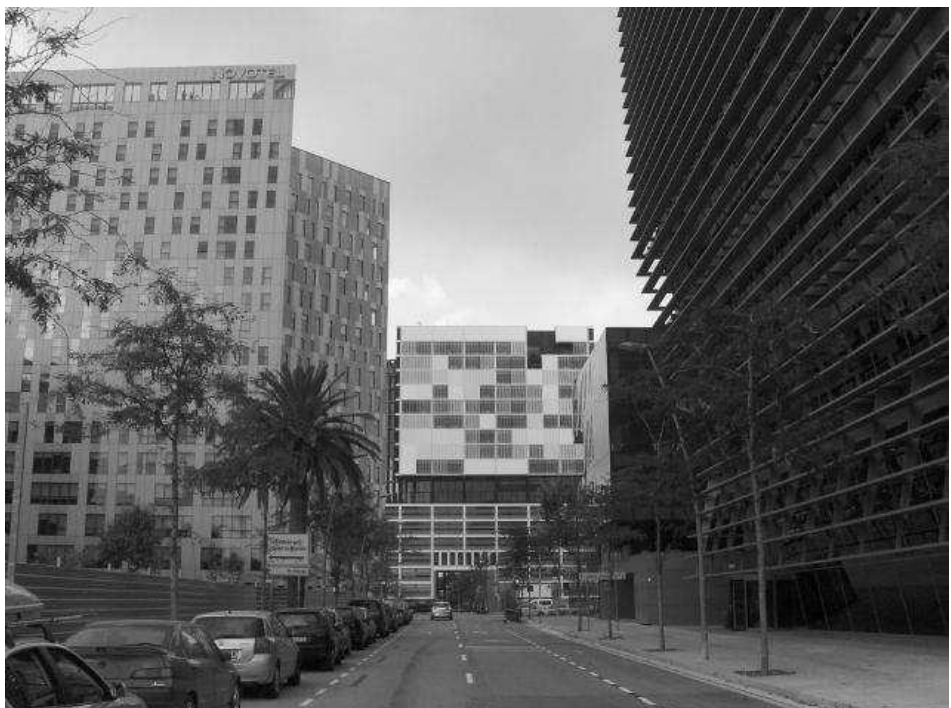

Un nouveau paysage urbain méditerranéen, 2010, quartier 22@, PB.

### 3.2 Entre dépendances et iniquités sociales chroniques : les exemples d'actions symboliques mémorielles patrimoniales

Durant cette période, la lutte pour le *«sauvetage»* de Can Ricart ne naît pas du hasard des convulsions d'un quartier ; on peut la rapporter aux rythmes des temporalités des fermetures d'ateliers d'artistes dans la nouvelle zone 22@ grâce à un travail de récolte de données. Une véritable montée en puissance des actions collectives se produit avec des articles militants sur Can Ricart, son sauvetage et même des contre-projets d'architecture et d'aménagements fort bien documentés et visibles sur le web, accompagnés de séries d'expositions publiques ; c'est la classe des intellectuelles qui prend position : architectes, professeurs de géographie et artistes de renommée viennent en aide aux occupants et à leurs revendications (Montaner, J.M. & Muxí, Z., 2006).
Un programme de couverture médiatique est planifié pour faire prendre conscience des dégâts de l'urbanisme 22@ avantageant les entreprises et non les habitants (M. Martí, 2008). Douze mois d'occupation de l'usine, suivis d'une expulsion mouvementée et de la fermeture du squat, obligent la mairie à proposer à son tour un Plan stratégique de la Culture via l'Institut Culturel de

Barcelone[4]. En toute logique, l'année 2006 permet, par la législation, de caractériser Can Ricart de bien d'intérêt culturel patrimonial (Mairie de Barcelone, 2006). Ses tensions urbaines montrent le manque de concertation ou d'obligation dans le partage foncier (Sabaté, Tironi, 2008). Les publicités de promoteurs célèbrent un quartier à la mode, mais niant le fait que Poblenou reste un vaste chantier, zone de non-droit pour habitants squatters, lieu de boîtes de nuits bruyantes et surtout zone de la politique innovante et modèle pour une équité sociale qui n'a réalisé que 7 % des logements sociaux prévus. Car les artistes sont aussi les premiers à revendiquer des logements pour tous en centre-ville. En 2010, 576 logements sociaux sur les 40 000 attendus horizon 2020 sont réalisés, ceci à rapporter au total de 42 619 logements construits entre 2000 et 2010, L'Espagne en pleine crise hypothécaire a stoppé le rythme des constructions à Poblenou, mais aussi leur vente, sur les 8556 habitations construites entre 2009 et 2011, 5 504 n'ont toujours pas trouvé acquéreur en 2011.

**3.3 La réalité des interconnexions entre citoyens et industries créatives**

Actuellement, que ce soit sur la Rambla de Poblenou, dans les squats des réfugiés africains ou les nouveaux restaurants du quartier, les habitants se questionnent et débattent des avantages de ce projet et de la réalité sur la mixité sociale comme enjeu de la qualité de vie. Des exemples comme *la mémoire de nos ancêtres*, proposent par le biais du numérique de faire se rencontrer des jeunes artistes créateurs et personnes âgées pour raconter et exposer l'héritage symbolique et associatif du quartier, de même avec le *22@CreaTalent* où ici les firmes (et uniquement elles) proposent une initiation aux outils informatiques pour les primaires. Derniers exemples, le *computer recycling* permet aux entreprises de 22@ de donner des ordinateurs à la population et à quelques artistes. Les jeunes de Poblenou bénéficient de cette implantation avec 500 étudiants qui ont pu réaliser un stage dans ces entreprises ; de même le réseau 22@network est une opération sur le web permettant de mettre en lien ces entreprises et parfois sur le plan de l'intranet de privilégier l'embauche de locaux.
Néanmoins, on ne peut nier que certains mouvements ou associations de quartiers, contre associations et comités de sauvegarde sont parfois surmédiatisés et surreprésentés dans les médias du fait de leur importance politique et d'un supposé poids auprès de l'électorat. La gentrification ou l'arrivée massive de populations de travail, de logements et d'un nouveau paysage urbain couvert de gratte-ciel, est certes parfois difficile à admettre pour des extrémistes politiques - n'appartenant et n'habitant pas dans les quartiers bien souvent (!) ; en revanche, elle ne l'est pas pour les propriétaires qui ont vu leur patrimoine revalorisé. Les locataires sont toujours dans cette histoire et cette géographie de quartier les plus mal lotis comme les studios d'artistes qui sont passés de plus de 200 en 2002 à moins d'une vingtaine actuellement (Narrero, 2003), un renouveau est à noter depuis peu.

---

[4] <http://w3.bcn.es/V01/Home/V01HomeLinkPl/0,2460,7610_52522_3,00.html> au 01/07/10

Enfin, les services publics connaissent une refonte totale dont l'accès pour handicapé et les stations de métro en cours de rénovation suite au retour négatif de la gestion des transports au sein du quartier.

## CONCLUSION

La société catalane du XXIe siècle a produit un espace néo-capitaliste fragmenté et dominateur par une frénésie immobilière irréfléchie et des jeux fonciers complexes. En parcourant les rues de Poblenou, on est obligé de penser aux travaux d'Henri Lefebvre sur la production de l'espace et les trois ordres de compréhension de la conception d'une composition urbaine (Lefebvre, 1974). Dans ce cas précis, l'unité entre l'espace physique, l'espace mental et l'espace social n'est que théorique et imparfaite dès le départ de sa conceptualisation. Si chaque société doit produire son espace, Poblenou est celui de la spéculation dont la véritable valeur resurgit dans les épreuves et luttes urbaines protestant contre une sorte de modélisation forcée d'un espace urbain par rapport à des modèles non européens grâce à la technique du benchmarketing. Néanmoins, la créativité est une notion qui semble être au cœur des problématiques de durabilité territoriale, elle permet, en effet, d'insuffler une dynamique et une réponse active des êtres humains face à des potentialités innovantes : *«la créativité peut être une réponse aux problématiques de durabilité territoriale »* (Lazzeri, 2010). La réussite du projet 22@ appartiendra aux concepteurs capables d'adapter le quartier à sa capacité de logements et de créations d'emplois pérennes tout en répondant aux exigences d'une population de plus en plus revendicative au sein de la cité catalane. Ne léser aucune catégorie sociale reste un défi dans une cité où le tourisme a pris le pas sur l'originalité ; touristes qui parcourent depuis peu le quartier à la recherche de sa nouvelle identité ou de son ancienne comme dans les visites du cimetière de Poblenou. Ici, la ville est devenue un enjeu et le lieu de la croissance économique reposant sur l'innovation. Enfin, la vie de quartier est en complète mutation ce qui paraît pour ce territoire laisser longtemps en marge, une chance.

## RÉFÉRENCES

Acebillo, J, 2004, Une nueva línea de horizonte, *BMM*, Barcelone, AB
Andreu, M., Gol, J.Y, Recio, A, 2007, "22@, acara o creu", *El Poblenou,* n°21, Barcelone, DB
Archive historique de Poblenou, 1998, *Gent de Poblenou*, Barcelone, AHP
Barceló, M., 2001, *Pacte Industrial de la Regió Metropolitana, La Ciutat Digital*, Barcelona, Beta Editorial
Arroba s.a.u., 2010, *10 anys de 22@:el districte de la innovació,* Barcelona: Ayuntamiento de Barcelona-Arroba s.a.u.
Barceló, M., 2010, «Faire entrer Barcelone dans le réseau mondial des villes créatives», *Barcelone, la ville innovante*, Paris : Le Moniteur
Borja, J., 2004, *Urbanismo en el siglo XXI. Bilbao, Madrid, Valencia, Barcelona,* Barcelone, Enta

Broggi, A., 2007, «Le projet 22@ Barcelone, mutation d'un espace industriel ou creuset d'innovation ?», *Cahiers IUARIF*, n°146

Busquets, J., 2004, *Barcelona, la construcción urbanística de una ciudad compacta*, Barcelone, Ediciones del Serbal

Checa, M., 2002, *Poblenou: la fàbrica de Barcelona*, Barcelone, Mairie de Barcelone

El Poblenou, 2001, *El Poblenou: més de 150 anys d'historia*, Barcelone, Arxiu històric del Poblenou

Florida, R., 2002, *The rise of the creative class*, New York, Basic Books

Florida, R., 2005, *Cities and the creative class*, New York, Routledge

Lefebvre, H, 1974, *La production de l'espace*, Paris, Anthropos

Lefevre, C., Romera, M, (Dir.), 2007, «Barcelone : la ville comme facteur de développement économique », *Synthèse du groupe de travail «Les modalités de mise en œuvre des stratégies de développement économique par des métropoles européennes*, Paris : IAURIF

Laigle, L, 2009, *Vers des villes durables*, Paris, PUCA

Lazzeri, Y.,2010, *Créativité, industries créatives, territoires créatifs: un état de l'art,* PDDTM, les notes du Pôle, UPC A-M III

Mairie de Barcelone, 1998, *Criteris, objectius i solucions generals de planejament de la renovació de les àrees industrials del Poblenou*. Barcelone,AB

Mairie de Barcelone, 2000, *Modificación del PGM para la Renovación de las Áreas Industriales del Poblenou – Distrito de Actividades* Barcelone Mairie de Barcelone, 2003, *Pla Estratègic Metropolità de Barcelona*. Barcelone, AB

Mairie de Barcelone, 2006, Modificació del Pla especial de protecció del Patrimoni arquitectònic historicartístic de la ciutat de Barcelona, Disctrict de Sant Martí, Patrimoni industrial del Poblenou, Barcelone, AB

Marti, M, 2008, El proyecto 22@bcn: glocal governance, renovación urbana y lucha vecinal en Barcelona, *Actas Democracia y buen Gobierno*, Barcelone

Martin, A, Pallares, A, Tulla, A, 2008, *Nueva economía y los espacios industriales tradicionales: el caso del 22@ Barcelona*, Barcelone, Département de Géographie, UB

Masboungi, A. (Dir.), 1998, *Barcelone, la seconde renaissance*, Paris, Projet Urbain DGUHC

Masboungi, A. (Dir.), 2010, *Barcelone, la ville innovante*, Paris, Projet Urbain, Le moniteur

Montaner, J. M., 2006, «Le modèle Barcelone», *La pensée de midi,* n°18, Marseille : Des Sud/Actes Sud

Moustier, E., Lazzeri, Y., 2006, « Bilan des expériences territoriales dans l'élaboration d'indicateurs de développement durable » dans *Les indicateurs territoriaux du développement durable*, Lazzeri, Y., (Dir.), Paris, L'Harmattan

Montaner, J.M. & Muxí, Z., 2006, "Primeres reflexions per teixir fàbrica, nous edificis i parc. Can Ricart,l'alterproposta" in Claros, S., et al. (ed.) *Can Ricart. Patrimoni, Innovació I Ciutadania.* Grup de Patrimoni Industrial del Fòrum de la Ribera del Besòs, Barcelona.

Narrero, I., 2003, *¿Del Manchester catalán al Soho Barcelonés? La renovación del barrio del Poble Nou en Barcelona y la cuestión de la vivienda*. Scripta Nova. Revista electrónica de geografía y ciencias sociales. Barcelone, Universidad de Barcelona, vol. VII, núm. 146

Sabate, J., Tironi, Y.M., 2008, «Rankings, creatividad y urbanismo», *Eure,* n° 102, Vol. XXXIV, Santiago de Chili, PUSC

Solà-Moralés, M., 1974, *Barcelona. Remodelación capitalista o desarrollo urbano en el sector de la Ribera Oriental*, Barcelone, Gustavo Gili

Oliva, A, 2003, *El districte d'activitats 22@bcn*, Barcelone, Fundació Bosch Gimpera, Universitat de Barcelona Colección Aula Abierta, Model Barcelona Quaderns de Gestió,nº 15

Pacte Industriel de la région Métropolitaine, 2001, *La Ciutat Digital*. Barcelona, Beta

Porter, M.E., 1998, «Clusters and the new economics of competition», *Harvard Business Review*, n°98 vol. 76 issue 6, Harvard: Harvard Business School Publishing

Tugaye, Z., 2009, *Réhabilitation de friches industrielles: 22@Barcelona*, Paris, feuillet de Eureka21.eu

Yigitcanlar, T., Velibeyoglu, K., Baum, S., 2008, *Creative urban regions: harnessing urban technologies to support knowledge city initiatives*, NY, IGI Global snippet

Walliser, A., 2006, «A place in the world: Barcelona's quest to become a global knowledge city», *Built Environment*, Oxon: Alexandrine press, 30

Site internet :

Association 22@, www.22network.net, 01/01/11
District 22@, http://www.22barcelona.com, 01/01/11
Mairie de Barcelone, service de l'urbanisme, www.bcn.es/22@,01/01/11
La Vanguardia, www.lavanguardia.es/afondo/vivienda, 01/01/11

**AUTEUR**

**Patrice BALLESTER**
patriceballester@yahoo.fr
Ater à l'Université de Pau et des Pays de l'Adour
Chercheur associé au lab. Geode UMR 5602

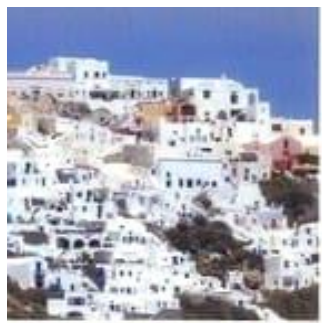

# VULNÉRABILITÉ, ÉQUITÉ ET CRÉATIVITÉ EN MÉDITERRANÉE

**ESPACE & DÉVELOPPEMENT DURABLE**

Cette collection est un lieu de rencontre où chercheurs et praticiens du développement durable partagent leurs savoirs, outils et pratiques.

Le développement durable conduit à de nouveaux enjeux pour les territoires et les politiques publiques, en matière de consommation, de déplacement, d'organisation urbaine etc. et renouvelle les modes de faire, avec des approches plus intégrées, plus transversales et plus participatives. Parler de développement durable impose de faire le point sur les menaces qui pèsent sur les territoires, mais plus encore cela impose de trouver des solutions innovantes, qui en appellent à l'imaginaire et à l'imagination.

C'est l'objet de cet ouvrage que de proposer des formes originales de recherche sur le développement durable, autour des notions de vulnérabilité, d'équité et de créativité territoriale.

En couverture
Ville de Fira,
Île de Santorin, Grèce
© Yvette Lazzeri

**Yvette Lazzeri** est chercheur au Centre d'études et de recherches internationales et communautaires d'Aix-Marseille Université (CERIC). Elle est responsable du pôle Développement durable et territoires méditerranéens (PDDTM).

**Emmanuelle Moustier** est maître de conférences, chercheur au Centre d'études et de recherche en gestion d'Aix-Marseille (CERGAM) d'Aix-Marseille Université, et membre du PDDTM.

Aix+Marseille université | Presses Universitaires de Provence | Presses Universitaires d'Aix-Marseille

9 782853 998550

24 €